\documentclass[electronic]{vgtc}                          

\ifpdf
  \pdfoutput=1\relax                   
  \usepackage{graphicx}                
  \DeclareGraphicsExtensions{.pdf,.png,.jpg,.jpeg} 
\else
  \usepackage{graphicx}                
  \DeclareGraphicsExtensions{.eps}     
\fi%

\graphicspath{{figures/}{pictures/}{images/}{./}} 

\usepackage{microtype}                 
\PassOptionsToPackage{warn}{textcomp}  
\usepackage{textcomp}                  
\usepackage{mathptmx}                  
\usepackage{times}                     
\usepackage{cite}                      
\usepackage{tabu}                      
\usepackage{booktabs}                  

\usepackage{array}

\usepackage{balance}
\usepackage{siunitx} 
\usepackage{amsfonts} 
\usepackage{amsmath} 
\usepackage{multirow} 

\onlineid{1102}

\vgtccategory{Poster}

\vgtcinsertpkg

\title{Remote Assistance with Mixed Reality for Procedural Tasks}

\author{Manuel Rebol\thanks{e-mail: mrebol@american.edu}\\ %
    \scriptsize American University %
             \and Colton Hood\thanks{e-mail: chood@mfa.gwu.edu}\\ %
 \scriptsize George Washington University
          \and Claudia Ranniger\thanks{e-mail: cranniger@mfa.gwu.edu}\\ %
 \scriptsize George Washington University
          \and Adam Rutenberg\thanks{e-mail: arutenberg@mfa.gwu.edu}\\ %
 \scriptsize George Washington University
          \and  Neal Sikka\thanks{e-mail: nsikka@mfa.gwu.edu}\\ %
 \scriptsize George Washington University
          \and Erin Maria Horan\thanks{e-mail: ehoran@american.edu}\\ %
 \scriptsize American University
         \and Christian Gütl\thanks{e-mail: c.guetl@tugraz.at}\\ %
 \scriptsize Graz University of Technology
      \and Krzysztof Pietroszek\thanks{e-mail: pietrosz@american.edu}\\ %
     \scriptsize American University 
}

\abstract{
We present a volumetric communication system that is designed for remote assistance of procedural tasks. The system allows a remote expert to visually guide a local operator. The two parties share a view that is spatially identical, but for the local operator it is of the object on which they operate, while for the remote expert, the object is presented as a mixed reality ``hologram''. Guidance is provided by voice, gestures, and annotations performed directly on the object of interest or its hologram. At each end of the communication, spatial is visualized using mixed-reality glasses.
} 

\keywords{Remote Procedure, Augmented Reality, Virtual Presence.}

\CCScatlist{
  \CCScatTwelve{Human-centered computing}{Interaction design}{}{};
  \CCScatTwelve{Computing methodologies}{Mixed / augmented reality}{}{}
}

\begin{document}



\maketitle

\section{Introduction}

Remote collaboration is playing an increasingly important role in people's everyday lives. With more people working remotely, video communication has become a \emph{de facto} standard for remote collaboration. Yet, in some applications, due to a lack of spacial information, a limited field of view, a lack of context, and a limited transmission of non-verbal cues and body language, the affordances provided by video communication are insufficient. One example where traditional video communication is not sufficient is remote assistance in procedural tasks. 

To address the deficiencies of video communication, previous work has explored domain specific systems for remote assistance using virtual, augmented, or mixed reality. Examples include remote assistance during space flight emergency \cite{datcu2014virtual}, remote surgery \cite{wisotzky2019interactive}, robotics operation\cite{gurevich2012teleadvisor,ArInRobotics2017}, telehealth \cite{ARtelemedicine2017}, and many others. The proposed remote assistance systems show promise, but are typically expensive and highly-specialized with their domains of use. 

In this work, we introduce a general-purpose volumetric communication system that can be used for remote assistance across many domains. To increase the accessibility, we develop our communication system using off-the-shelf hardware. While in our example application, we focus on remotely guided assembly procedures, our system can be deployed for other remote communications situations in which a remote expert guides a local operator. Currently, our system supports two users in a one-to-one communication scenario, but we plan to support one-to-many and many-to-many communication scenarios. 
Compared to previous domain specific systems, our volumetric communication system consists of widely available off-the-shelf hardware. Moreover, our system is developed such that it can be used for different applications without hardware changes. We focus on high portability to allow for fast deployment whenever a remote procedure needs to be conducted. To provide an intuitive collaborative experience, we virtually place the remote expert and the local operator side-by-side similar to how they would interact if the remote expert was physically present.

\section{Mixed-Reality Remote Assistance}

Our system supports audiovisual, spatially-enhanced communication and interaction between two parties, the local operator and the remote expert. The local operator works on a procedural task while being assisted by a remote expert providing audio and visual guidance. 

\begin{figure}
    \centering
    \includegraphics[width=\linewidth]{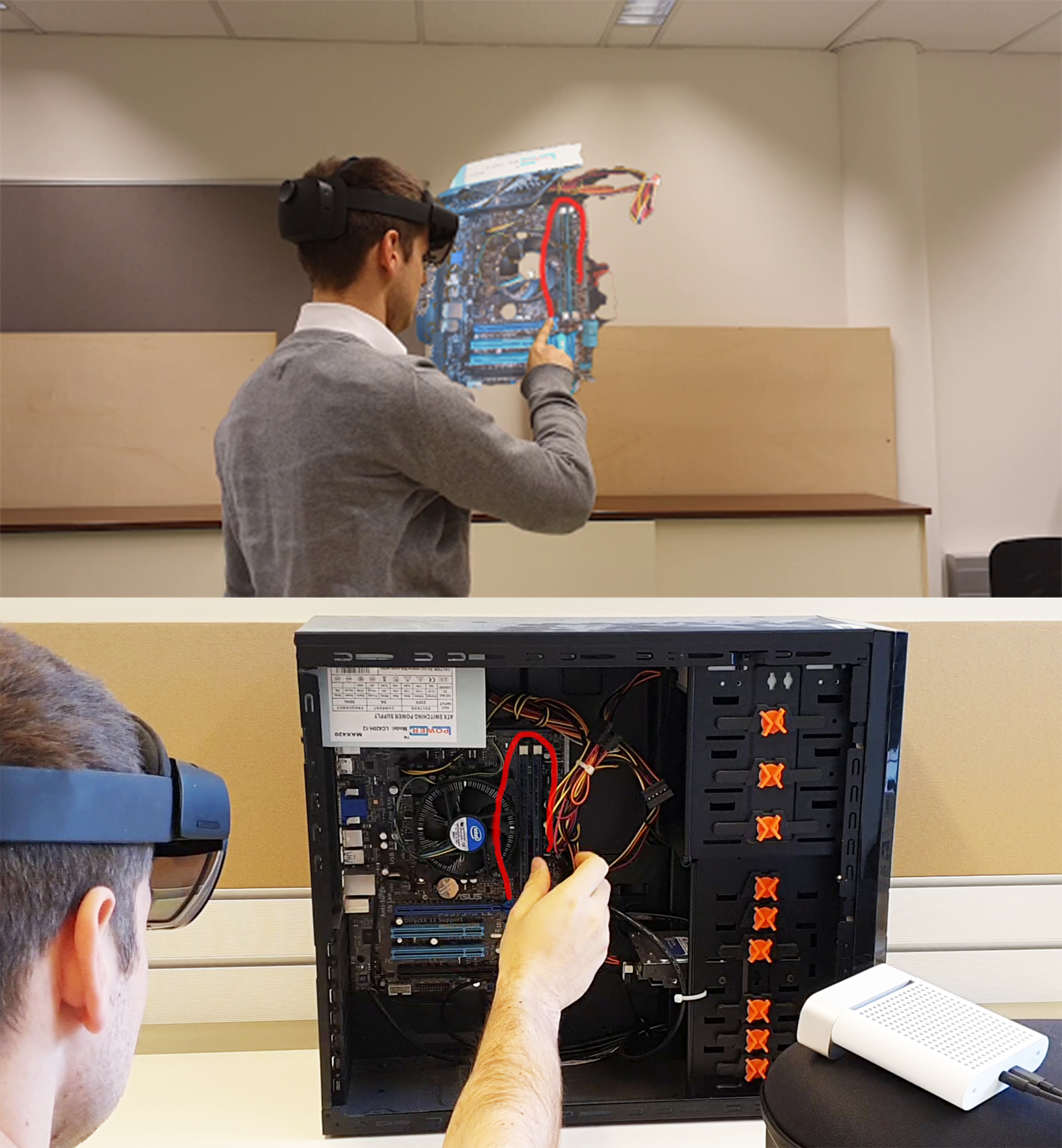}
    \caption{Top: Remote expert view, including annotations. Bottom: Local operator's view. The expert is adding annotations which are visible to local operator.}
    \label{fig:annotations}
\end{figure}

\paragraph{Remote Expert View}
When using traditional video communication, the essential problem the remote expert is facing, when training or assisting local operators, is insufficient spatial information. Thus, the main functionality of our system is to provide spatial information in real time. To do so, our system continuously scans the object of interest, tracks the position of the hands and tools used by the local operator, and presents this information as a holographic visualization to the remote expert. The ``holographic'' rendering of the object of interest is as realistic as a two-dimensional video recording, but it additionally includes the shape of the object (\autoref{fig:annotations}). Importantly, spatial sound of the local operator's environment is also communicated to the remote expert, increasing immersion of the expert in the local operator's environment. 

Using the spatially-rendered view of the local operator's object of interest and its context, the remote expert visually guides the local operator in performing a procedure. The expert uses speech, pointing and annotating topology of the object of interest. 

An example of the view of the expert is shown in \autoref{fig:annotations}. As illustrated, the remote expert can annotate the view and that annotation is visible to the local operator. In our example scenario, the procedure is to replace a part in desktop computer, so the remote expert annotates the slot in the motherboard where the replacement part should be placed.

\paragraph{Local Operator View}
Since the local operator is present at the physical place of execution of the procedure, he/she sees a physical object of interest in front of themselves, and therefore does not need to be presented with the ``hologram'' of the object of interest. However, the spatial information about the shape of the object of interest is used on the local operator's side to enable displaying remote expert's annotation directly on the physical object of interest. Additionally, the local operator can annotate the physical object of interest and these annotations are visible to the expert, providing two way annotation-based communication between the local and the remote party. 

\subsection{Hardware}
In our system, both users are equipped with a lightweight, standalone, high-resolution mixed-reality head-mounted display. The head mounted display is used to present, in real-time, a rendering of three-dimensional data \emph{registered} to real objects in the users' environment. For our prototyping, we choose the Hololens 2, but in principle any mixed-reality headset capable of freehand interaction could be used.

The local operator's system is equipped with volumetric capture camera that continuously feeds high-fidelity spatial data to the remote expert. We use the volumetric capture of the low-cost time-of-flight depth camera Kinect v4 Azure with a separate processing unit. In our setup, the volumetric capture is performed from the position where a supervising operator would have position themselves, if they were co-located with the local operator. 
Our choice of hardware is informed by a desire to increase accessibility of spatial communication. We aim to keep the overall system cost as low as possible, as compared to the existing industry-standard telepresence systems. The total cost of our system with the above described components is around \$5,000 per user. 

\subsection{Implementation}
At the local side, the scene is captured with the Kinect Azure (Kinect v4) volumetric camera. The color information is retrieved by the RGB camera and the depth by the time-of-flight sensor. We process both data streams using the Azure Kinect SDK. Once the streams are processed, they are forwarded to the Mixed Reality WebRTC \cite{webRTCvideo} libraries that handle the peer-to-peer connection between the remote and local operator, avoiding most of the issues with firewalls and NAT traversals. The camera sends both depth and color data in real time over network to the remote expert. The camera sends only the depth data to local operator. The color frames are compressed using VP9 
video compression codec. To save on data bandwidth, for depth data we compare frames over time and only send those parts of the depth that have changed.   

The WebRTC client running on the head mounted display of the remote and local operator receives the depth streams. Additionally, the remote expert receives the color stream. For the remote view, we assemble the depth and color information and render a 3D mesh. As a result, the local scene is visualized with high spatial fidelity on the Hololens 2 head mounted display of the remote expert.

Besides viewing the local scene, the expert also gives verbal and visual instructions to the local operator. Using their hands, he/she can provide pointing directives and annotations. We track the hand with the kinect camera built in the Hololens 2. We use the  Mixed-Reality Toolkit 
to recognize the hand gestures, delineating between pointing and annotation gestures. The gestures and the annotations are sent to the Hololens 2 of the local operator. The annotations and gestures provided by local operators are relayed to the remote expert in the same manner.

We are able to keep the end-to-end latency of the system below 500 milliseconds. In our pilot studies, the delay is not noticeable, because no instant interaction between the operators takes place on the visual level. 

\subsection{Evaluation}
The feasibility of the volumetric system prototype has been tested in a pilot study within our research group. Initial implementation suffered from over 3 seconds latency. However, by introducing above-mentioned compression and switching to WebRTC for network communication, this insufficiency has been resolved. We plan to perform a full user study for the system in medical remote assistance in a near future, when the Covid-19 pandemic subsides allowing us back on campus.

\section{Conclusion}
We present a real-time volumetric communication system designed for remote assistance. Our system allows a distant remote expert to guide a local operator on a procedural task through virtual gestures and annotations. The guidance is given using mixed-reality headsets which support intuitive and efficient communication. Our system is low cost and can be used in various industries including healthcare, logistics and education. 

\acknowledgments{
The work is supported by National Science Foundation grant no. 2026505 and 2026568. We would also like to thank Ilia Kowalzik for their help in developing the system.
}


\bibliographystyle{abbrv-doi}

\bibliography{references}
\end{document}